\documentclass{ifacconf}

\usepackage{graphicx}      
\usepackage{amsmath}
\usepackage{float}
\usepackage{amssymb}
\usepackage{stfloats}
\usepackage{siunitx,booktabs,array}
\usepackage{mathtools}
\usepackage[backend=biber,style=authoryear,date=year,doi=false,url=false,isbn=false,maxcitenames=2,maxbibnames=99,giveninits=true,uniquename=false,uniquelist=false]{biblatex}
\begin{document}
\begin{frontmatter}

\title{Delayed dynamic-feedback controller design for multi-frequency vibration suppression} 

\thanks[footnoteinfo]{The presented research was supported by the project G092721N of the Research Foundation-Flanders (FWO-Vlaanderen) and by the Czech Science Foundation under the project 24-10301S. The research was also co-funded by European Union under the project Robotics and Advanced Industrial Production No. $CZ.02.01.01/00/22\_008/0004590$. The second author acknowledges support by the Grant Agency of the Czech Technical University in Prague, grant No. SGS24/125/OHK2/3T/12.\\ \\
\textsf{\large Preprint submitted to IFAC}}

\author[KUL,CTU]{Adrian Saldanha} 
\author[CTU]{Adam Peichl} 
\author[KUL]{Wim Michiels}
\author[CTU]{Tomas Vyhlidal}

\address[KUL]{Department of Computer Science, KU Leuven, Celestijnenlaan~200A, 3001~Leuven, Belgium (\{adrian.saldanha,wim.michiels\}@kuleuven.be)}
\address[CTU]{Dept. Instrumentation and Control Engineering, Faculty of Mechanical Engineering and Czech Institute of Informatics, Robotics and Cybernetics, Czech Technical University in Prague, Prague~6, Czechia ({adam.peichl,tomas.vyhlidal}@fs.cvut.cz)}

\begin{abstract}                
We present a methodology for designing a dynamic controller with delayed output feedback for achieving non-collocated vibration suppression with a focus on the multi-frequency case. 
To synthesize the delay-based controller, we first remodel the system of equations as a delay-differential algebraic equation (DDAE) in such a way that existing tools for design of a static output feedback controller can be easily adapted. The problem of achieving non-collocated vibration suppression with sufficient damping is formulated as a constrained optimization problem of minimizing the spectral abscissa in the presence of zero-location constraints, with the constraints exhibiting polynomial dependence on its parameters. We transform the problem into an unconstrained one using elimination, following which we solve the resulting non-convex, non-smooth optimization problem.
\end{abstract}

\begin{keyword}
vibration suppression, time-delay, DDAE
\end{keyword}

\end{frontmatter}


\section{Introduction}
\label{sec:1}


In the context of vibration suppression, the setting wherein a so-called absorber or actuating device is placed at a location remote from the target point of suppression is referred to as the \emph{non-collocated} setting. The need for such a setting usually follows from the practical impossibility of placing an absorber precisely at the target location. This makes controller design a challenging task, as any actuation effort for suppressing disturbances at the remote target must be transmitted to the target via the interconnected masses.

Noteworthy studies on the topic of non-collocated vibration suppression include: \textcite{buhrNonCollocatedAdaptivePassive1997}, \textcite{yangTimeDelayNoncolocated1992} and even more recently, \textcite{olgacActivelyTunedNoncollocated2021}. Most of these works target the single-frequency setting. In \textcite{silmSpectralDesignExperimental2024}, the authors propose a control design method involving direct assignment of imaginary zeros of the transfer function constructed from the harmonic disturbance to the target, applicable to the multi-frequency setting as well. However, the design necessitates the use of two separate controllers for expanding the admissible frequencies.

We approach the multi-frequency case in a similar way i.e. first, we recast the problem of suppressing vibrations of finite discrete frequencies into assigning transmission zeros of the transfer function, constructed from the external force to the target mass's position, on the imaginary axis at the selected frequencies. One can find vast literature on similar \emph{anti-resonance assignment} techniques (see \textcite{richiedeiUnitrankOutputFeedback2022}, \textcite{richiedeiPolezeroAssignmentReceptance2022}). We refer the text \textcite{richiedeiActiveApproachesVibration2021} for a detailed survey of literature on such methods. 
In the non-collocated problem setting, the number of outputs available for feedback are often limited, and therefore the available free controller parameters are insufficient. This poses a limitation on the number of frequencies that can be targeted, or makes it difficult to achieve a higher-level control task such as stabilization, tracking control, etc. 

In \textcite{saldanhaStabilityOptimizationTimedelay2024}, the author's team presented a non-collocated controller design methodology using a single control input and alternately with multiple inputs. The admissible frequencies in the single-input setting were limited by the number of outputs available for feedback. For the multi-input setting instead, the greater number of free parameters resulting from the extra inputs provide for more admissible frequencies. However, this is limited by the number of actuators physically available and in the ideal setting, we are interested in developing a control methodology using a single-input, which can be expandable to the multi-input setting. In \cite{saldanhaNoncollocatedVibrationSuppression2023}, the authors propose a design method using a single control input, by adding multiple delays in the feedback to maximize the stability margin, but this method was applied to a single frequency setting.

We expand on this idea of adding multiple delays in order to extend the range of admissible frequencies. The addition of multiple delays to the output feedback can be seen as incorporating derivative information and fits into the framework of intentionally adding delays for control purposes \parencite{abdallahDelayedPositiveFeedback1993}, \parencite{niculescuStabilizingChainIntegrators2004}. The classic delayed resonator \parencite{olgacTunableActiveVibration1995} is an ideal example of using delays in vibration suppression. Valasek et al. (\cite{valasekRealtimeTunableSingledegree2019}) extended the scope of the delayed resonator control scheme to handle multiple and time-varying frequencies, using a single-mass absorber. The control scheme utilizes position feedback and multiple delays to target a total of two discrete frequencies, and it applies to the collocated setting only.

In this work, we target the \emph{multi-frequency} vibration suppression problem. The solution can be seen as an extension of the methodology presented in the aforementioned article of \textcite{saldanhaNoncollocatedVibrationSuppression2023}. The considered controller design bears a nice property in that the zero-location constraints were affine in the controller parameters $K$. In this text, in addition to delaying the outputs by multiple fixed delays, we feed the delayed outputs to a dynamic feedback controller of fixed order. With the presence of the dynamic feedback, it is shown later that the constraints are no longer affine in these parameters, but instead are polynomial in these parameters. We demonstrate how to remodel the problem in terms of delay-differential algebraic equations (DDAEs), enabling a formulation of the design problem as a constrained optimization problem in a form which closely resembles that of the multi-input setting in \textcite{saldanhaOptimizationbasedAlgorithmSimultaneous2022}. 

The rest of the paper is organized as follows. In Section \ref{sec:2}, we outline the problem and provide a motivation for the adopted controller structure, following which, we present the control design methodology in Section \ref{sec:3}. The application of the proposed method is demonstrated in a case study in Section \ref{sec:4}, with the conclusions in Section \ref{sec:5}.

\section{Motivation and Problem Formulation}
\label{sec:2}

\subsection{Problem setting}
\label{subsec:problem_setting}

\begin{figure}[h]
    \centering
    \includegraphics{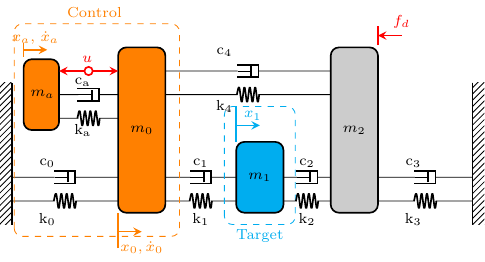}
    \caption{The figure shows a minimal example of a non-collocated setup. The control goal here is to tune the control input $u$ to cancel vibrations at the target $m_1$, subject to an external harmonic disturbance $f_d$ acting at the mass $m_2$.}
    \label{fig:smd-system}
\end{figure}

Let us assume a vibration controller setup modelled by the equations
\begin{equation}\label{eq:1}
    \begin{aligned}
    \dot{x}(t)      &= A_0 x(t) + B_1 u(t-\tau_u) + B_2 f_d(t) \\
    y(t)            &= C_1 x(t) \\
    z(t)            &= C_2 x(t), \\
    \end{aligned}
\end{equation}
where $x$ is the state vector, $y$ the feedback outputs and $z$ representing the position of the target where we wish to cancel vibrations. The matrix $A_0$ is the state matrix corresponding to the non-delayed state $x(t)$. $B_2\in \mathbb{R}^{n\times 1}$ describes the entry point corresponding to the external disturbance force $f_d(t) = \sum\limits_{k=1}^m F_d \cos (\omega_k t + \phi)$ with $\omega_k = 2\pi f_k$, $f_k$ being the discrete harmonic frequencies and $C_2 \in \mathbb{R}^{1\times n}$ describes the target mass's position. Control input matrix $B_1$ corresponds to control input $u$, which is subject to inherent time-delay $\tau_u$. Feedback output matrix $C_1$ describes measured outputs $y$ available to the controller. In the general setting $B_1 \in \mathbb{R}^{n\times n_u}$ and $C_1\in\mathbb{R}^{n_y \times n}$ i.e. we consider the setting with multiple inputs and outputs, with $n_u$ and $n_y$ representing the number of control inputs and measured outputs respectively. 
A minimal example of a non-collocated setup is shown in Figure \ref{fig:smd-system}. Here, the disturbance $f_d$ acts on the mass $m_2$, thereby exciting the remaining masses. The control goal here is to cancel vibrations at the target mass $m_1$ by tuning the control input $u$, supplied by a voice-coil actuator, located between the masses $m_a$ and $m_0$.

\subsection{Control objective}
We focus on designing a \emph{low-authority controller} (LAC) for suppressing vibrations of multiple harmonics $\omega_k, k = 1,\ldots,m$. In addition to the objective of vibration suppression, we also focus on maximizing the stability margin of the resulting closed-loop system. 

To motivate the adopted controller, let us first look at the case of a static output feedback controller without delays i.e. 
\begin{equation}\label{eq:2}
    u(t) = K^T y(t).
\end{equation}
Without the loss of generality, we formulate our methodology for a single-input controller, which can easily be expanded to a controller with multiple inputs. For the single-input setting $B_1\in \mathbb{R}^{n\times 1}$ and with $n_y$ outputs, we have $C_1 \in \mathbb{R}^{n_y\times n}$. The gain matrix $K \in \mathbb{R}^{n_y}$, with which, we have $n_y$ free parameters available for manipulation.

The objective of suppressing vibrations of a set of frequencies can be translated into a problem of placing the transmission zeros of the transfer function from the external force $f_d$ to the position of the target mass $z$. Cancelling single frequency oscillations requires at least two controller parameters and for $m$ harmonics, we would  require a minimum of $2m$ free controller parameters. In the limiting case, $n_y=2m$, no free parameters are available for any additional task such as tracking or stability optimization.



Instead, we present a two-step complementary method of extending the scope of a static output feedback controller:
\begin{enumerate}
    \item using intentional output delays and
    \item using dynamic output feedback with the delayed output measurements
\end{enumerate}

Starting from the output vector $y$, we first delay this output by multiple fixed delays viz. $\tau_i, i=1,\ldots,N$ and create a new vector of delayed inputs $y_d$ as
\begin{equation}\label{eq:3}
    y_d(t)          = [y_1^T \quad y_2^T \quad \ldots \quad y_N^T]^T
\end{equation}
where $y_i = y(t-\tau_i), \quad i = 1, \ldots, N$, $N$ being the number of delays. Here, $y_d \in \mathbb{R}^{n_y N}$. Due to the $N$ added delays, we now have $n_yN$ free parameters available for control.

The presence of an input delay in itself results in an infinite-dimensional closed-loop system. The infinite-dimensionality of the system further motivates for the addition of more delays in the output.

\begin{figure}[h]
    \centering
    \includegraphics{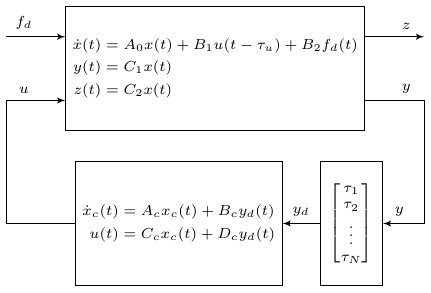}
    \caption{Block diagram of the dynamic controller with delayed feedback}
    \label{fig:delayed-feedback}
\end{figure}

In the next step of the design, the delayed outputs are fed to a dynamic feedback controller of the form
\begin{equation}\label{eq:4}
    \begin{aligned}
        \dot{x}_c(t)    &=      A_c x_c(t) + B_c y_d(t)   \\
        u(t)            &=      C_c x_c(t) + D_c y_d(t).   \\
    \end{aligned}
\end{equation}
Here, $A_c \in \mathbb{R}^{n_c\times n_c}$, $B_c \in \mathbb{R}^{n_c\times ({n_y N})}$, $C_c \in \mathbb{R}^{n_u\times n_c}$ and $D_c \in \mathbb{R}^{n_u\times ({n_y N})}$.
Figure \ref{fig:delayed-feedback} shows the schematic diagram of the closed-loop system with the delayed feedback and the dynamic controller part.

\section{Methodology}\label{sec:3}


The inherent delays in the input and the artificially induced delays in the feedback outputs, together with the presence of dynamic feedback increase the design complexity. To account for this added complexity, the system of equations is remodelled as a Delay-Differential Algebraic Equation (DDAE). We use DDAEs since it helps to account for various feedback interconnections using a relatively simple structure. Additionally, for our controller design, we use the \texttt{MATLAB}\textsuperscript{\textregistered} package \texttt{TDS-CONTROL}, which is a package for designing controllers for time-delay systems that supports DDAE models.

\subsection{Remodelling the system}
\label{subsec:reform_ddae}

By introducing a new state vector $\Tilde{x} = [x^T \ \zeta_u^T \ x_c^T \ \zeta_y^T]^T$, we can rewrite Equations \eqref{eq:1}, \eqref{eq:2} and \eqref{eq:4} as a DDAE:

\begin{equation}\label{eq:5}
\begin{aligned}
    \Tilde{E} \Dot{\Tilde{x}}(t) &= \begin{multlined}[t]
    \Tilde{A}_0 \Tilde{x}(t) + \Tilde{A}_u \Tilde{x}(t-\tau_u) + \Sigma_{i=1}^{N} \Tilde{A}_i \Tilde{x}(t-\tau_i) \\ + \Tilde{B}_2 w(t) + \Tilde{B}_1 \Tilde{u}(t) \\
    \end{multlined} \\ \vspace{-10pt}
    z(t)        &= \Tilde{C}_2 \Tilde{x}(t) \\
    \Tilde{y}(t)        &= \Tilde{C}_1 \Tilde{x}(t) \\
\end{aligned}
\end{equation}
coupled with a controller of the form
\begin{equation}\label{eq:6}
    \Tilde{u}(t)       = K \Tilde{y}(t),
\end{equation}
with $\zeta_u^T$, $\zeta_y^T$ denoting slack variables corresponding to the control input and the delayed measured outputs respectively and $K$ can be partitioned as $K = \left[ \begin{array}{c | c}
     A_C    &   B_C \\
     \hline
     C_C    &   D_C \\
\end{array} \right]$, $K \in \mathbb{R}^{(n_c+n_u)\times (n_c+n_y N)}$. We denote the number of elements of the gain matrix $K$ as $n_p$, with $n_p = (n_c+n_u)\times (n_c+n_y N)$.

The resulting DDAE in Equation \eqref{eq:5} represents a retarded time-delay system. Note also that with the remodeled system, the Equation \eqref{eq:6} now resembles one of a static output feedback controller. For more details on moving input-output delays to state delays and modelling dynamic feedback in the DDAE framework we refer to \textcite{gumussoyFixedOrderHInfinityControl2011}.



\subsection{Assigning a pair of transmission zeros}
\label{subsec:trans_zeros}

Suppose that the harmonic disturbance force $f_d$ enters into the system remotely. This causes the system to oscillate at the same frequencies. Suppressing vibrations of a single discrete frequency at a target point corresponds to a pair of transmission zeros of the transfer function constructed from the external force to the target mass's position being located on the imaginary axis at the specific excitation frequency. 
Formally, a transmission zero i.e. $\lambda_z=\pm j \omega$ can be characterized by the solution to:
\begin{equation}\label{eq:7}
\scriptsize 
    \begin{vmatrix}
        j\omega \Tilde{E} - \Tilde{A}_0 - \Tilde{A}_u e^{-j\omega \tau_u} - \sum_{i=1}^{N} \Tilde{A}_i e^{-j\omega \tau_i} - \Tilde{B}_1 K \Tilde{C}_1  & -\Tilde{B}        \\
        \Tilde{C}                   & 0         \\
    \end{vmatrix} = 0.
\end{equation}
To fully suppress the multi-harmonic disturbances $f_d$, we constrain the controller parameters such that the closed-loop zeros are located at $\pm j\omega_i, \ i = 1,\ldots,m$.

These constraints can be formalized as follows:
\begin{equation}\label{eq:8}
    h_k(K) = 0,    \quad k = 1,\ldots,m
\end{equation}
with  
\scriptsize 
\begin{multline*}
    h_k(K) = \\    
    \begin{vmatrix}
        j\omega_k \Tilde{E} - \Tilde{A}_0 - \Tilde{A}_u e^{-j\omega_k \tau_u} - \sum_{i=1}^{N} \Tilde{A}_i e^{-j\omega_k \tau_i} - \Tilde{B}_1 K \Tilde{C}_1  & -\Tilde{B}        \\
        \Tilde{C}                   & 0         \\
    \end{vmatrix}
\end{multline*}
\normalsize
specifying the zero-location constraints.

\subsection{Stability optimization}
\label{subsec:stabopt}


The characteristic matrix for a time-delay system described by Equations \eqref{eq:5} and \eqref{eq:6} is given by:
\begin{multline}\label{eq:9}
    M(\lambda;K) = \lambda \Tilde{E} - \Tilde{A}_0 - \Tilde{A}_u e^{-\lambda \tau_u} - \sum_{i=1}^{N} \Tilde{A}_i e^{-\lambda \tau_i}  - \Tilde{B}_1 K^T \Tilde{C}_1
\end{multline}

The delay in the input, coupled with the added delays to the output feedback leads to an infinite-dimensional system with infinitely many characteristic roots. 
The spectral abscissa function, in turn can be defined as the supremum of the real parts of the characteristic roots of the closed-loop by:
\begin{equation}\label{eq:10}
    \alpha(K) := \sup \left(\Re(\lambda): |M(\lambda;K)|=0 \right).
\end{equation}

Maximizing the stability margin of the overall system corresponds to minimizing the spectral abscissa function. 
This requirement of minimizing the spectral abscissa function $\alpha$ in the presence of zero-location constraints as defined in \eqref{eq:8} can be modelled as an optimization problem of the form:

\begin{equation}\label{eq:11}
    \begin{aligned}
        \min_{K}  \quad \alpha(K) \\
        \text{subject to:} \quad h_k(K) &= 0, \quad k=1,\ldots,m.
    \end{aligned}
\end{equation}

The optimization problem is in general non-smooth and non-convex \parencite{michielsStabilityStabilizationTimedelay2007}. Additionally, we see that the constraints in Equation \eqref{eq:10} are polynomial in its parameters $K$. The infinite-dimensionality of the system together with the polynomial constraints make the problem challenging to solve directly using available software.

\subsection{Constraint Elimination}

We use constraint elimination to solve the ensuing optimization problem.  Specifically, we apply the property that the constraint functions $h_i$ are affine in a subset of its controller parameters $K$. Using this property, we split the gain $K$ into $n_p-2m$ \textit{independent} parameters and $2m$ \textit{dependent} parameters, where the latter are used to assure constraint satisfaction. The dependent parameters are selected from a single row or a single column of the $K$ to ensure the affinity property. We refer to the procedure described in \textcite{saldanhaOptimizationbasedAlgorithmSimultaneous2022} for details about the choice of dependent and independent parameters.

According to the procedure from \textcite{saldanhaStabilityOptimizationTimedelay2024}, we separate the term $\Tilde{B}_1 K \Tilde{C}_1$ from Equation \eqref{eq:8} into its \textit{dependent} and \textit{independent} parts:
\begin{equation}\label{eq:12}
    \Tilde{B}_1 K \Tilde{C}_1 = \Tilde{B}_1 K_L \Tilde{C}_1 + \Tilde{B}_{11} g^T \Tilde{C}_{g}
\end{equation}
where $\Tilde{B}_{11}$ indicates the input matrix corresponding to one input, $g$ the  dependent gains and $\Tilde{C}_{g}$ the output matrix corresponding to the dependent outputs.
Here, the vector $g$ represents the dependent variables while $K_L$ is the matrix which consists of the independent parameters to be optimized. The elements of $K_L$ corresponding to the dependent parameters are set to $0$ while the rest are free parameters for optimization.

By placing a pair of zeros at $\lambda_k = \pm j \omega_k $ and by applying the Weinstein-Aragozajn identity, we reformulate the constraints from Equation \eqref{eq:11} as:
\begin{equation}\label{eq:13}
     1 - g^T [\Tilde{C}_{g} \quad 0] R(\omega_k,K_L)^{-1} \begin{bmatrix}
        \Tilde{B}_{11} \\ 0
    \end{bmatrix}
     = 0
\end{equation}
assuming $R$ is invertible, where 
\begin{multline}\label{eq:14}
\scriptsize
    R(\omega_k,K_L) = j\omega_k \begin{bmatrix}
        \Tilde{E}   & 0 \\
        0           & 0
    \end{bmatrix} - 
    \begin{bmatrix}
        \Tilde{A}_0   & \Tilde{B} \\
        -\Tilde{C}  & 0
    \end{bmatrix} - \begin{bmatrix}
        \Tilde{A}_u   &     0 \\
        0       &   0
    \end{bmatrix} e^{-j\omega_k \tau_u}
     \\ 
     \scriptsize - 
    \sum_{i=1}^N
    \begin{bmatrix}
        \Tilde{A}_i   &     0 \\
        0       &   0
    \end{bmatrix} e^{-j\omega_k \tau_i} - 
    \begin{bmatrix}
        \Tilde{B}_1 K_L^T \Tilde{C}_1       & 0 \\
        0                                   & 0
    \end{bmatrix}.
\end{multline}
Hence, we obtain a set of equations of the form:
\begin{equation}\label{eq:15}
    \begin{aligned}
        \Re\{ g^T z(\omega_k,K_L)\}     &=  1    \\
        \Im\{ g^T z(\omega_k,K_L)\}     &= 0    \\
    \end{aligned}
\end{equation}
where $z(\omega_k,K_L) = \begin{bmatrix}
    \Tilde{C}_g & 0
\end{bmatrix} R(\omega_k,K_L)^{-1} \begin{bmatrix}
    \Tilde{B}_{11} \\ 0
\end{bmatrix}$.
For $m$ pairs of zeros, we get $2m$ separate equations in $g$ as:
\begin{equation}\label{eq:16}
    P(K_L) g = Q  \iff g(K_L) = P(K_L)^{-1} Q
\end{equation}
provided the matrix $P(K_L)$ is invertible.

\subsection{Unconstrained optimization problem}
\label{subsec:unconstr}
With the dependent parameters $g$ eliminated by Equation \eqref{eq:15}, we obtain an unconstrained optimization problem in $K_L$:
\begin{equation}\label{eq:17}
    \begin{aligned}
        \min_{K_L}  \quad \hat{\alpha}(K_L) \\
        \hat{\alpha}(K_L) &:= \sup \left(\Re(\lambda): |M_1(\lambda;K_L)|=0 \right)
    \end{aligned}
\end{equation}
with \begin{multline*}
M_1(\lambda;K_L) = \lambda \Tilde{E} - \Tilde{A}_0 - \Tilde{A_u} e^{-\lambda \tau_u}   \\ - \sum_{i=1}^{N} \Tilde{A}_i e^{-\lambda \tau_i} - \Tilde{B}_{11} g^T(K_L) \Tilde{C}_{g} - \Tilde{B}_{1} K_L \Tilde{C}_{1}
\end{multline*}

The optimization problem in Equation \eqref{eq:16} is nonsmooth and nonconvex, although, it is differential almost everywhere. We utilize the BFGS algorithm with weak Wolfe line search from the MATLAB\textsuperscript{\textregistered} software package \texttt{HANSO} (Hybrid solver for nonsmooth optimization) \parencite{lewisNonsmoothOptimizationQuasiNewton2013} together with the \texttt{TDS-CONTROL} \parencite{appeltansAnalysisControllerdesignTimedelay2023} package for solving the ensuing problem. The solver relies on computing the objective function and its derivative, wherever it exists.

\section{Case Study}
\label{sec:4}

The controller design is validated by simulation on a minimal spring-mass-damper system, shown in Figure \ref{fig:smd-system}. The system consists of four masses $m_a$, $m_0$, $m_1$ and $m_2$. An external harmonic disturbance $f_d=\sum\limits_{k=1}^m F_d \cos \omega_k t$ with $F_d = 3 N$ and $\omega_k = 2\pi f_k, \quad k = 1,\ldots,4$, and $f_k = 4,8,12,16$ Hz enters into the system via the mass $m_2$ which in turn causes the remaining masses to oscillate at the same frequencies. Here the target is $m_1$ which is not directly accessible, in that it is not possible to place an actuator directly onto this mass and secondly because it is assumed to be impossible to physically measure the position or velocity of this mass. The control effort $u$ is instead supplied by means of a voice-coil actuator located between the masses $m_a$ and $m_0$. The inherent input-delay $\tau_u$ is measured as $2 \text{ ms}$. The goal is then to parameterize $u$ to completely annihilate oscillations at the mass $m_1$.
Here we consider feedback measurements of positions and velocities of the masses $m_a$ and $m_0$ only (highlighted in orange in Figure \ref{fig:smd-system}).

The parameters of the system are shown in Table \ref{tab:1}.
\newcolumntype{M}{>{\centering\arraybackslash$}p{1.4cm}<{$}}
\begin{table}
  \caption{Parameter values}
  \centering \setlength{\tabcolsep}{0.0cm}
  \begin{tabular}{M S[table-column-width=1.4cm]M S[table-format=4.0,table-column-width=1.4cm]M S[table-column-width=1.4cm]}
    \toprule
    \multicolumn{2}{c}{Mass in \si{kg}} & \multicolumn{2}{c}{Stiffness in \si{Nm^{-1}}} & \multicolumn{2}{c}{Damping in \si{Nsm^{-1}}} \\ \midrule
    m_a & 0.52      & k_a   & 407   & c_a   & 1.8 \\
    m_0 & 1.1750    & k_0   & 1001  & c_0   & 4.35   \\
    m_1 & 0.5050    & k_1   & 749   & c_1   & 0.85   \\
    m_2 & 0.7290    & k_2   & 711   & c_2   & 1.85   \\
        &           & k_3   & 950   & c_3   & 4.95   \\
        &           & k_4   & 377   & c_4   & 0   \\
    \bottomrule
  \end{tabular}
  \label{tab:1}
\end{table}
We first consider a linear state-space model of the system in the form of Equation \eqref{eq:1}. Here, the state $   x = \begin{bmatrix}
        x_0     & \Dot{x}_0 &   x_1     & \Dot{x}_1 & x_2   & \Dot{x}_2     & x_a   & \Dot{x}_a
    \end{bmatrix}^T$, system matrix $A$ in Table \ref{tab:State Matrix}, the measured output $y = \begin{bmatrix}
        x_0     & \Dot{x}_0     & x_a   & \Dot{x}_a
    \end{bmatrix}^T$ with \[C_1 = \begin{bmatrix}
        1 & 0 & 0 & 0 & 0 & 0 & 0 & 0 \\
        0 & 1 & 0 & 0 & 0 & 0 & 0 & 0 \\
        0 & 0 & 0 & 0 & 0 & 0 & 1 & 0 \\
        0 & 0 & 0 & 0 & 0 & 0 & 0 & 1 
    \end{bmatrix}\] and the control output $z = x_1$ with $C_2 = \begin{bmatrix}
        0 & 0 & 1 & 0 & 0 & 0 & 0 & 0 
    \end{bmatrix}$. The input matrices corresponding to the control input and the external disturbance are $B_1 ~= \begin{bmatrix}
        0 & -\frac{1}{m_0} & 0 & 0 & 0 & 0 & 0 & \frac{1}{m_a} 
    \end{bmatrix}^T$ and $B_2 = \begin{bmatrix}
    0 & 0 & 0 & 0 & 0 & \frac{1}{m_2} & 0 & 0
\end{bmatrix}^T$ respectively. The system matrix $A$ is in Table \ref{tab:State Matrix}.
\begin{table*}[tp] \label{tab:State Matrix}
\rule{\linewidth}{\heavyrulewidth}
\[A = \begin{bmatrix}
    0 & 1 & 0 & 0 & 0 & 0 & 0 & 0 \\
    -\frac{k_0+k_1+k_a+k_4}{m_0} & -\frac{c_0+c_1+c_a+c_4}{m_0} & \frac{k_1}{m_0} & \frac{c_1}{m_0} & \frac{k_4}{m_0} & \frac{c_4}{m_0} & \frac{k_a}{m_0} & \frac{c_a}{m_0} \\
    0 & 0 & 0 & 1 & 0 & 0 & 0 & 0 \\
    \frac{k_1}{m_1} & \frac{c_1}{m_1} & -\frac{k_1+k_2}{m_1} & -\frac{c_1+c_2}{m_1} & \frac{k_2}{m_1} & \frac{c_2}{m_1} & 0 & 0 \\
    0 & 0 & 0 & 0 & 0 & 1 & 0 & 0 \\
    \frac{k_4}{m_2} & \frac{c_4}{m_2} & \frac{k_2}{m_2} & \frac{c_2}{m_2} & -\frac{k_2+k_3+k_4}{m_2} & -\frac{c_2+c_3+c_4}{m_2} & 0 & 0 \\
    0 & 0 & 0 & 0 & 0 & 0 & 0 & 1 \\
    \frac{k_a}{m_a} & \frac{c_a}{m_a} & 0 & 0 & 0 & 0 & -\frac{k_a}{m_a} & - \frac{c_a}{m_a}
\end{bmatrix}
\]
\end{table*}

 \begin{figure}
    \centering
    \includegraphics{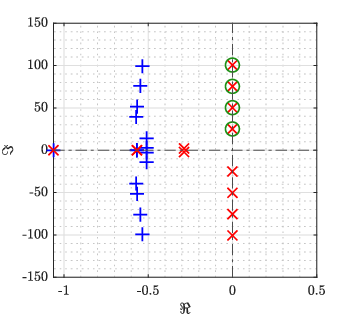}
    \caption{Pole-zero plot showing the rightmost eigenvalues of the closed-loop system using static controller with four delays. After tuning, the desired zeros are assigned correctly and stability margin is sufficient.}
    \label{fig:spectra}
\end{figure}

\begin{figure*}[ht]
    \centering
  \includegraphics[scale=0.85]{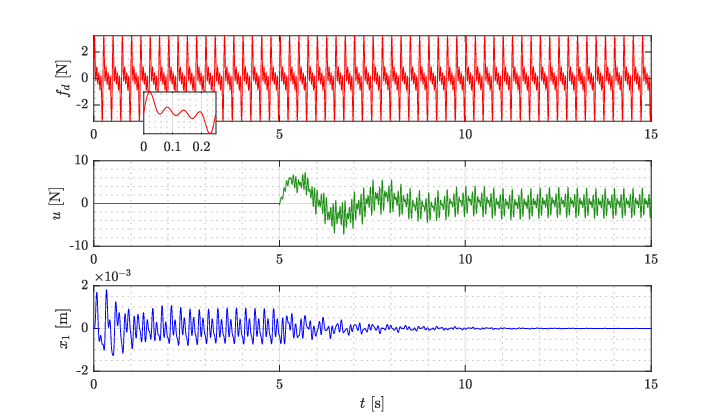}
  \caption{Multi-harmonic disturbance force $f_d(t)$ (top) acting on mass $m_2$, controller output $u(t)$ (middle) and displacement $x_1(t)$ of target mass $m_1$ (bottom). Controller is turned on at time $t=5~s$, following which, after transient period, the oscillation of target mass is silenced.}
\end{figure*}

\subsection{Using static output feedback, no delays}
\label{subsec:static_output}
With four target frequencies, we would require at least eight free controller parameters, one for each zero. However, by using a simple static feedback controller, we are left with at most four free parameters (corresponding to the four measured outputs), which are insufficient for placing the eight zeros. We can increase the free parameters by adding at least two delays in the feedback, which in turn leads to a total of exactly eight free parameters for placing the eight zeros. This leaves us with no additional free parameters for optimizing the spectral abscissa as per Equation \eqref{eq:17}.


\subsection{Using static output feedback, with four delays}
\label{subsec:static_output_delayed}
As per the proposed controller design method, we add four equi-spaced delays in the interval  $(0.05, 0.2)$. The delay interval is chosen  based on the signal frequencies. The control input is parameterized according to the feedback law in Equations \eqref{eq:3} and \eqref{eq:4}. The addition of the four delays consequently leads to a four-fold increase in free parameters. We proceed to optimize the closed-loop spectrum subject to the prescribed zero-location constraints, using the procedure detailed in Section \ref{sec:3}. In the first step, the controller parameters are computed for a zero-order controller (a static one). Next, the controller order is increased by one in successive steps, with the controller parameters initialized with the computed lower-order controller setting. Consequently, the resulting optimum solution achieved with the higher-order controller can be no lower than the optimum computed for the previous lower-order controller. 

The results of the optimization are shown in Table \ref{tab:2} (the controller parameters are not shown in the table due to limited space). We see that with four delays, we achieve complete closed-loop stability with $\alpha<0$. By increasing the order of the controller, we can obtain an even better result viz. the resulting spectral abscissa is lower. 
\begin{table}[]
    \centering
    \begin{tabular}{c c c c}
        \toprule
            $n_c$   & $n_d$     &   $\tau$  &     $\alpha$ \\
        \midrule
            0       &   4       &   $[0.05 \quad 0.1  \quad 0.15 \quad
            0.2]^T$ &      $-0.5218$   \\
            1       &   4       &   $[0.05 \quad 0.1  \quad 0.15 \quad
            0.2]^T$ &    $-0.5218$   \\
            2       &   4       &   $[0.05 \quad 0.1  \quad 0.15 \quad
            0.2]^T$ &    $-0.5322$   \\
            3       &   4       &   $[0.05 \quad 0.1  \quad 0.15 \quad
            0.2]^T$ &    $-0.5347$   \\
        \bottomrule
    \end{tabular}
    \caption{Closed-loop spectral abscissa for different orders of the controller}
    \label{tab:2}
\end{table}


\section{Discussion}
\label{sec:5}
We presented a methodology for designing a delayed-dynamic feedback controller for achieving multi-frequency non-collocated vibration suppression. The controller design, by itself is challenging due to intentionally added delays and the dynamic feedback component. However, with the use of DDAEs we systematically remodel the system of equations. With the help of these techniques, we were able to suppress vibrations of a total of four discrete frequencies. The robustness of the approach and the experimental validation are topics for further research.



\printbibliography

                                                   








\end{document}